\documentclass[twocolumn,floatfix,amssymb,aps,superscriptaddress,letterpager]{revtex4}
\usepackage[colorlinks=true,urlcolor=blue,citecolor=blue,linkcolor=blue]{hyperref}
\usepackage{graphicx}
\usepackage{color}
\usepackage{epsfig}
\usepackage{amsmath}
\usepackage{amssymb}
\usepackage{mathrsfs}
\usepackage{bm}
\usepackage{subfigure}
\usepackage{float}
\usepackage{url}
\usepackage{enumerate}
\usepackage{braket}
\usepackage{comment}
\usepackage{multirow}
\usepackage{makecell}
\definecolor{violet}{rgb}{0.56,0.0,1.0}

\begin{document}

\hsize\textwidth\columnwidth\hsize\csname@twocolumnfalse\endcsname

\title{Tensor network simulation for the frustrated $J_1$-$J_2$ Ising model on the square lattice}

\author{Hong Li}
\email[]{157luoluoluo@ruc.edu.cn}
\affiliation{Department of Physics, Renmin University of China, Beijing 100872, China}

\author{Li-Ping Yang}
\email[]{liping2012@cqu.edu.cn}
\affiliation{Department of Physics, Chongqing University, Chongqing 401331, China}

\begin{abstract}
By using extensive tensor network calculations, we map out the phase diagram of the frustrated $J_1$-$J_2$ Ising model on the square lattice. 
In particular, we focus on the cases with controversy in the phase diagram, especially the stripe transition in the regime $g = |J_2/J_1|>\frac{1}{2}$, $(J_2>0,J_1<0)$. 
While recent studies claimed that the phase transition is of first order when $\frac{1}{2}<g<g^*$ (with the smallest $g^*$ being $0.67$), our simulations suggest that if there is such a first-order region, it is smaller than those found in earlier studies by other methods.
Combining with the analysis of critical properties, we provide evidence that the classical $J_1$-$J_2$ model evolves continuously from two decoupled Ising models ($g\to\infty$ with central charge $c = 1$) to a point belonging to the tricritical Ising universality class (with $c = 0.7$) as $g$ decreases to $g^*\simeq 0.54$.
\end{abstract}

\maketitle
\section{introduction}
Onsager provided the rigorous solution of the two-dimensional classical Ising model~\cite{1}, which greatly deepened and enriched the understanding of phase transitions. As a natural generalization of this model, the next-nearest-neighbor interaction $J_2$ can be added, whose phase diagram and nature of the phase transition are, however, still under debate~\cite{2,3,4,5,6,7,8,27,9,10,11,12,13,14,15,16,17,19,20,21,22,23,24,25,26,huyi,28,29,30}. There is an intuitive physical picture as follows: For the frustrated $J_1$-$J_2$ Ising model at low temperature, the system lies in the ferromagnetically ordered phase when $g = J_2/|J_1|<\frac{1}{2}$, corresponding to $Z_2$ symmetry breaking; for $g>\frac{1}{2}$, the system enters the stripe antiferromagnetic (SAFM) phase, also called the collinear antiferromagnetic phase with $Z_4$ symmetry breaking. 

Compared to the corresponding quantum model, theoretical analysis and numerical simulations are easier in the classical case, which nevertheless sheds light on some common properties of frustrated systems. In particular, the SAFM phase is of great interest, since a moderate external field induces an adjacent nematic phase~\cite{31,32}. In its quantum counterpart, such a nematic phase might be useful for understanding the mechanism of the high-$T_c$ superconductor~\cite{33,34,35}.

Some earlier works show that, the phase transition between two ordered phases and the paramagnetic phase (above a threshold temperature) is of continuous type. It is commonly believed that the phase transition belongs to the Ising universality class in the case of $g<\frac{1}{2}$, while it fits the weak universality in the case of $g>\frac{1}{2}$~\cite{2,3,4,5,6,7,8,27}. For example, some critical exponents depend on the strength of interaction and therefore vary with the coupling parameters~\cite{39}. 
 In conformal field theory (CFT), a phase transition with varying critical exponents is possible for a central charge $c\geq 1$. Extensive numerical calculations, based on the cluster-variation method (CVM) and Monte Carlo (MC) simulations, have shown that for $\frac{1}{2}<g<g^*$ ($g^*\simeq 1$), the phase transition is of first order~\cite{9,10,11,12,13,14,15}, while other simulations for some given $g<g^*$ tend toward the conclusion of weakly universal continuous phase transition~\cite{16,17,19,20}.

The upper limit $g^*$ keeps decreasing with improved numerical simulations. For smaller $g$ above $\frac{1}{2}$, the bigger system size is needed for a reliable MC simulation. The $g^*$ varied from $0.9$~\cite{21} to $0.67$~\cite{22, 24}, where the $J_1$-$J_2$ Ising model was numerically mapped to the Ashkin-Teller (AT) model which also exhibits $Z_4$ symmetry breaking. 
Jin {\it et al}. found a weak first-order phase transition at $g<g^*$ and the pseudo-first-order behavior at $g\gtrsim g^*$~\cite{22, 24}.
Monte Carlo simulations with larger system sizes strengthened their conclusion~\cite{23} and the recent cluster mean-field (CMF) method~\cite{26} also mentioned consistent results. Recently, a numerial transfer matrix study was carried out for extensive frustrated lattice models~\cite{huyi}. In Table~\ref{prworks} we summarize previous results of $g>\frac{1}{2}$ in the literature.

Reaching the conclusion of both a weak first-order phase transition and a continuous phase transition with pseudo-first-order behaviors, the finite-size effect is a notorious issue which MC and CVM simulations cannot avoid. In contrast, tensor network algorithms allow us to approach the thermodynamic limit. Recently, tensor network methods were applied to the $J_1$-$J_2$ Ising model~\cite{adil} and pin down the Berezinskii-Kosterlitz-Thouless phase transition in the classical clock model~\cite{SL}. Levin and Nave proposed the tensor renormalization group (TRG) approach~\cite{Levin}. The TRG approach also performs well in the spin glass model~\cite{wangc}. Combining the TRG and higher-order singular value decomposition (HOTRG)~\cite{46}, the phase transition temperature of the three-dimensional Ising model was determined to the seventh decimal place~\cite{wangshun}.

\begin{table*}[htbp]
\renewcommand\arraystretch{1.3}
\caption{Comparison of results in previous works. }
	\begin{tabular} { l|c|c}
		\hline
		\hline
                 \makecell[c]{References} & Methods & Results\\
		\hline
                \multicolumn{3}{c}{$g>1/2$ with weak universality}\\
		\hline
		 \cite{2} & renormalization & \\ 
		 \cite{3} & MC renormalization group & \\ 
		 \cite{4} & high-temperature series & \\
		 \cite{5,6,7}  & MC & \\
		 \cite{8} & low-temperature and series expansion & \\
                \hline
                 \cite{16} & short-time MC & \\
                 \cite{17} & MC & no first-order transition at $g=1$  \\
                 \cite{19,20} & partition function zeros and MC  & \\
		\hline
                \multicolumn{3}{c}{$1/2<g<g^*$ with first-order transition}\\
		\hline
                 \cite{9} & CVM & $g^*=1.144$\\
                 \cite{10} & CVM & $g^*=1.045$\\
                 \cite{11} & mean field & $g^*=1.2$\\
                 \cite{12} & CVM, MC & $g^*=1.14$ (CVM) and $g^*=1.35$ (MC)\\
                 \cite{13} & effective-field theory & $g^*=0.95$\\
                 \cite{14,15} & MC & $g^*\sim1$\\
                 \cite{21} & MC and field-theoretic methods & $g^*=0.9$ \\
                 \cite{22,24} & MC, CMF, and transfer-matrix methods & $g^*=0.67$ \\             
                 \cite{23} & MC & vanishing of first-order signals at $g=0.8$\\
                  \cite{25} & effective-field theory & $g^*\sim1$\\
                 \cite{26} & CMF & $g^*=0.66$ \\
                 \cite{huyi} & numerical transfer matrix & first-order signals at $g = 0.55, 0.6$\\
		\hline
		\hline
	\end{tabular}
\label{prworks}
\end{table*}

In this article, combining infinite time-evolving block decimation (iTEBD)~\cite{44, 45} and the HOTRG, we make an effort to map out the full phase diagram and discuss the nature of the phase transition.  We find that the line of the first-order phase transition may not exist or lie in a much narrower region than those claimed in earlier studies when $g>\frac{1}{2}$. As $g$ goes down to about $\frac{1}{2}$, the central charge of the model decreases from 1 (two decoupled Ising models) to about 0.7, which corresponds to the tricritical Ising (TCI) universality class~\cite{36}. This result is further strengthened by the calculation of the Klein bottle (KB) entropy~\cite{49}. 

The rest of this paper is organized as follows. In Sec.~\uppercase\expandafter{\romannumeral2} we introduce the model and the related formulation based on iTEBD and the HOTRG. We employ iTEBD to calculate the physical quantities. The HOTRG helps us extract the conformal data from the fixed-point tensor during the renormalization group (RG) flow. In addition, we extract the information of critical theories by using the Klein bottle entropy~\cite{49}. In Sec.~\uppercase\expandafter{\romannumeral3}, we demonstrate and discuss our numerical results, including the phase diagram and the critical properties. We summarize our results in Sec.~\uppercase\expandafter{\romannumeral4}. 

\section{model and methods}
The Hamiltonian of the $J_1$-$J_2$ Ising model is written as
\begin{equation}
H=J_1 \sum\limits_{\left\langle ij\right\rangle }\sigma_i\sigma_j+J_2 \sum\limits_{\left\langle \left\langle ij\right\rangle \right\rangle }\sigma_i\sigma_j, \label{eq:model}
\end{equation}
where the spin variables take the values $\sigma = \pm 1$, and $J_1$ snd $J_2$ are coupling constants, corresponding to ferromagnetic ($J_1<0$) nearest-neighbor interactions denoted by $\left\langle ij \right\rangle $ and antiferromagnetic ($J_2>0$)
next-nearest-neighbor interactions denoted by $\left\langle \left\langle ij \right\rangle \right\rangle $, respectively. For convenience, we set $J_1=-1$ and $J_2=g$ hereafter. 
\subsection{Calculation of physical quantities}
\begin{figure}[h]
	\centering
	\subfigure[]{
		\includegraphics[width=4.2cm]{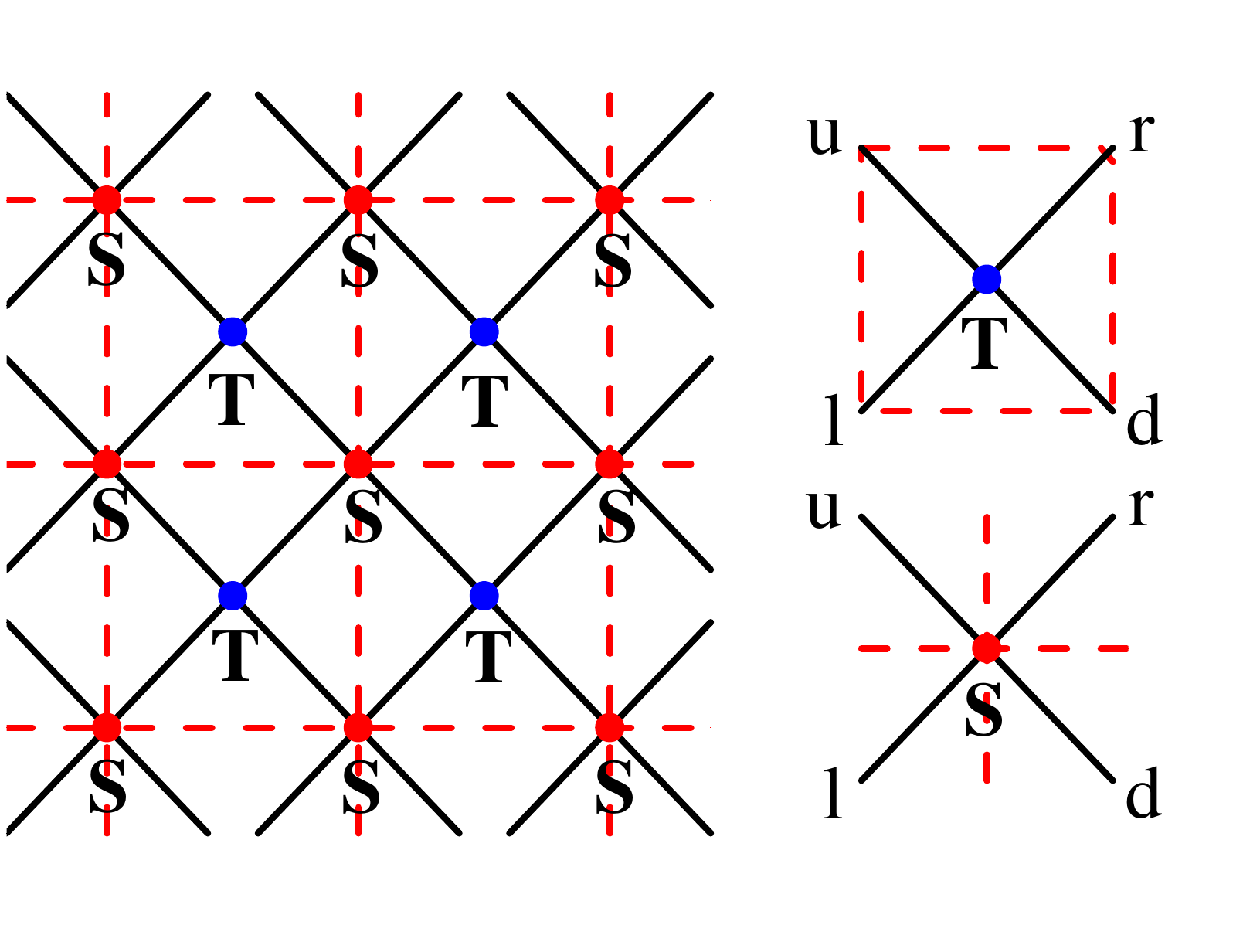}
	}
	\quad
	\hspace{-1cm}
	\subfigure[]{
		\includegraphics[width=4.2cm]{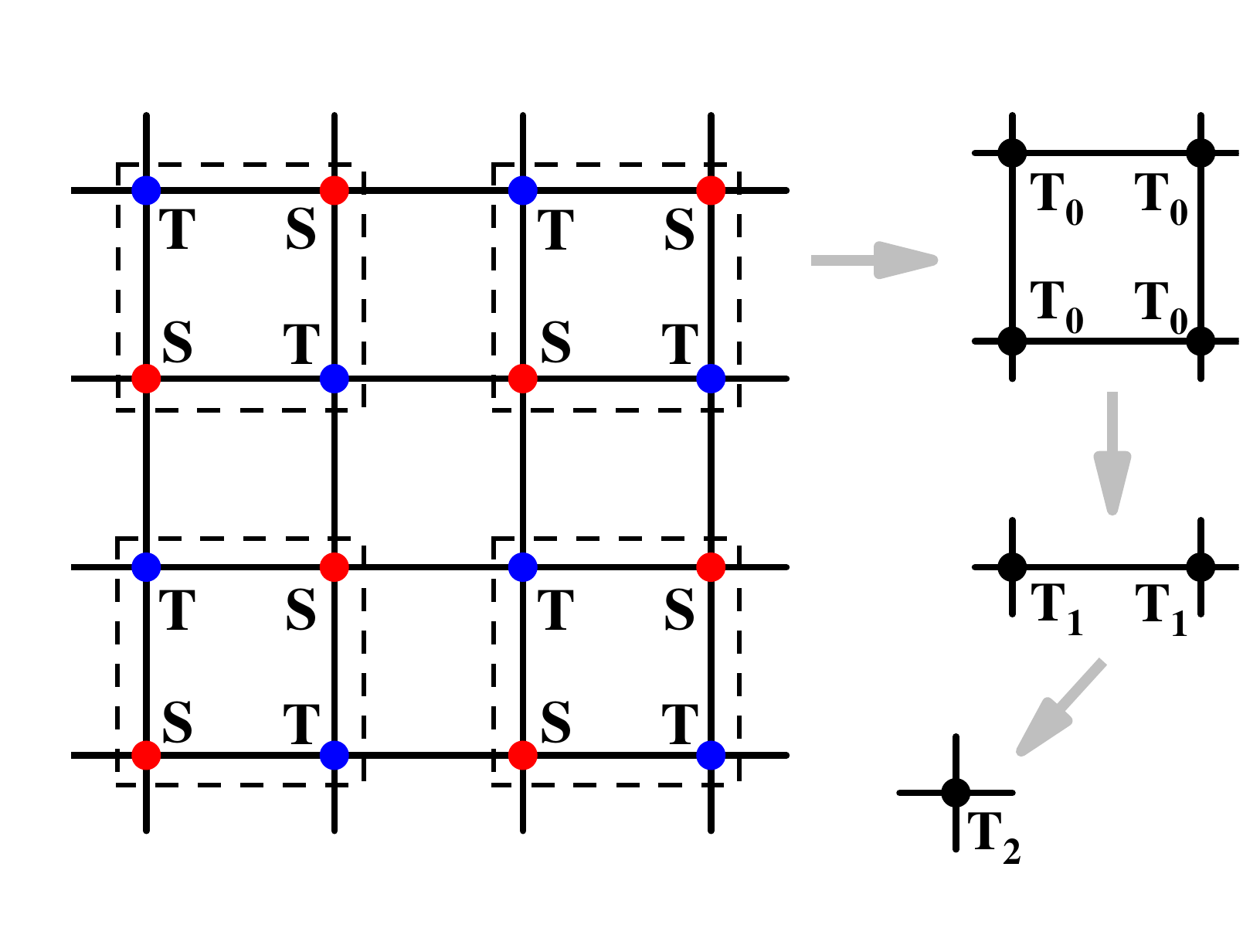}
	}
	\caption{(a) Defnition of tensors $T$ and $S$ with four indices $l,r,u~\text{and}~ d$. (b) Demonstration of coarse graining, where $T_0$ forms the single-site unit from the original $2\times2$ unit cell (enclosed by the dashed line).}\label{fig:TensorNcg}
\end{figure}

On the square lattice, the partition function of the model (\ref{eq:model}) can be cast into a tensor network form~\cite{huihai} 
\begin{equation}
Z=\sum_{\left\lbrace \sigma\right\rbrace} e^{-\beta H\left\lbrace \sigma\right\rbrace }=\text{Tr}(TSTS.~.~.),
\end{equation}
with the two types of tensor, $T$ and $S$, shown in Fig.~\ref{fig:TensorNcg}(a): $T$ is defined on the center of the square unit and $S$ is defined on the lattice site; $T$ and $S$ both hold the reflection symmetry in the $x$- and $y$- directions, whose explicit forms are given by
\begin{equation}
\begin{aligned}
&T_{lrud}=e^{(\beta/2)(\sigma_l\sigma_u+\sigma_u\sigma_r+\sigma_r\sigma_d+\sigma_d\sigma_l)-\beta g(\sigma_l\sigma_r+\sigma_u\sigma_d)},\\
&S_{lrud}=\delta_{lrud} =\left\{\begin{array}{lr}1~~\text{for}~l=r=u=d \\ 0~~\text{otherwise.}\end{array}\right.
\end{aligned}
\end{equation}
Here $\sigma_{l,r,u,d}$ refers to the four spin variables (left, right, up and down) as the tensor indices emitting from $T$.
As a consequence, we obtain a tensor network with a four-site unit cell [see Figs. 1(a) and (b)]; then, by coarse graining, $T_0$ is formulated as the single site tensor shown in Fig.~\ref{fig:TensorNcg}(b). Compared to the original lattice, the tensor network formed by $T_0$ is rotated by an angle $\pi/4$.

For the calculation of a local physical quantity $O$, we adopt the impurit site method~\cite{huihai}, in which only a single tensor $T$ encodes the operator $O$ (labeled $T^O$) and other tensors are unchanged. In this scheme, the internal energy $U$, the magnetization $m$, and the stripe magnetization $m_s=\sqrt{m_x^2+m_y^2}$ are defined as
follows:
\begin{equation}
\begin{aligned}
T_{lrud}^U=&-\frac{1}{2}[ (\sigma_l\sigma_u+\sigma_u\sigma_r+\sigma_r\sigma_d+\sigma_d\sigma_l)\\
&+g(\sigma_l\sigma_r+\sigma_u\sigma_d)]\times T_{lrud},\\
T_{lrud}^m=&\frac{1}{4}(\sigma_l+\sigma_r+\sigma_u+\sigma_d)\times T_{lrud},\\
T_{lrud}^{m_y}=&\frac{1}{4}(\sigma_l-\sigma_r-\sigma_u+\sigma_d)\times T_{lrud},\\
T_{lrud}^{m_x}=&\frac{1}{4}(\sigma_l-\sigma_r+\sigma_u-\sigma_d)\times T_{lrud}.
\end{aligned}
\end{equation}

The key point of iTEBD lies on the row-by-row projection based on the power method. The final result is a converged matrix product state (MPS).
The entanglement entropy of the converged MPS provides a useful way to locate phase transition points. By using the normalized entanglement spectrum $\left\lbrace \lambda_1,\lambda_2,..., \lambda_j, ...\right\rbrace $, with $\sum_j \lambda_j^2 = 1$,  the entanglement entropy reads
\begin{equation}
S_E=-\sum_{j}\lambda_j^2 \text{ln}(\lambda_j^2).
\end{equation}
Meanwhile, the correlation length can be calculated from the largest and the second largest eigenvalues $\epsilon _1$ and $\epsilon _2$ of the transfer matrix corresponding to the MPS as follows:
\begin{equation}
\xi=-\frac{1}{\text{ln}|\epsilon_2/\epsilon_1|}.
\end{equation}

\subsection{Extraction of critical properties}
In the framework of the HOTRG~\cite{46}, 
coarse graining is a renormalization group process, during which CFT information is encoded in the fixed-point tensor $T^{\ast}(=T^{(i)}/\sum_{lu}T^{(i)}_{lluu} )$, where $T^{(i)}$ is the tensor unit in the $i$th RG step and $T^{\ast}$ is unchanged under the RG process~\cite{48}.

Then we get $M_{u,d}=\sum_{l} T^{\ast}_{llud}$, with the eigenvalues: ${\Lambda_0, \Lambda_1,..., \Lambda_m, ...}$ (in descending order), from which
the central charge $c$ and scaling dimensions $h_m$ can be read~\cite{48}:
\begin{equation}
\begin{aligned}
&c=\frac{6}{\pi}\text{ln}\Lambda_0,\\
&h_m=-\frac{1}{2\pi}\text{ln}(\Lambda_m/\Lambda_0).\label{eq:cri}
\end{aligned}
\end{equation}

There is also an alternative way to obtain the central charge $c$ from the partition functions~\cite{49,affleck,HJM}. When the system is defined on the torus and the Klein bottle, the logarithms of the partition partitions are given by
\begin{equation}
\begin{aligned}
	&\text{ln}Z^T\simeq-f_0L_xL_y+\frac{\pi c}{6L_y}L_x,\\
	&\text{ln}Z^K\simeq-f_0L_xL_y+\frac{\pi c}{24L_y}L_x+S_\text{KB}, \label{eq:lnZTK}
\end{aligned}
\end{equation}
where $L_x$ and $L_y$ are the system sizes, $f_0$ is a nonuniversal constant, and $S_\text{KB}$ is known as the Klein bottle entropy~\cite{50}, which is universal. We note that the validity of Eq. (\ref{eq:lnZTK}) requires $L_x \gg L_y$. The iTEBD method allows us to calculate both torus and Klein bottle partition functions, so that the Klein bottle entropy $S_{\mathrm{KB}}$ and the central charge $c$ can be extracted as well.

\section{numerical results}
\subsection{Phase diagram and physical quantities}
\begin{figure}[h]
	\centering
	\includegraphics[scale=0.28]{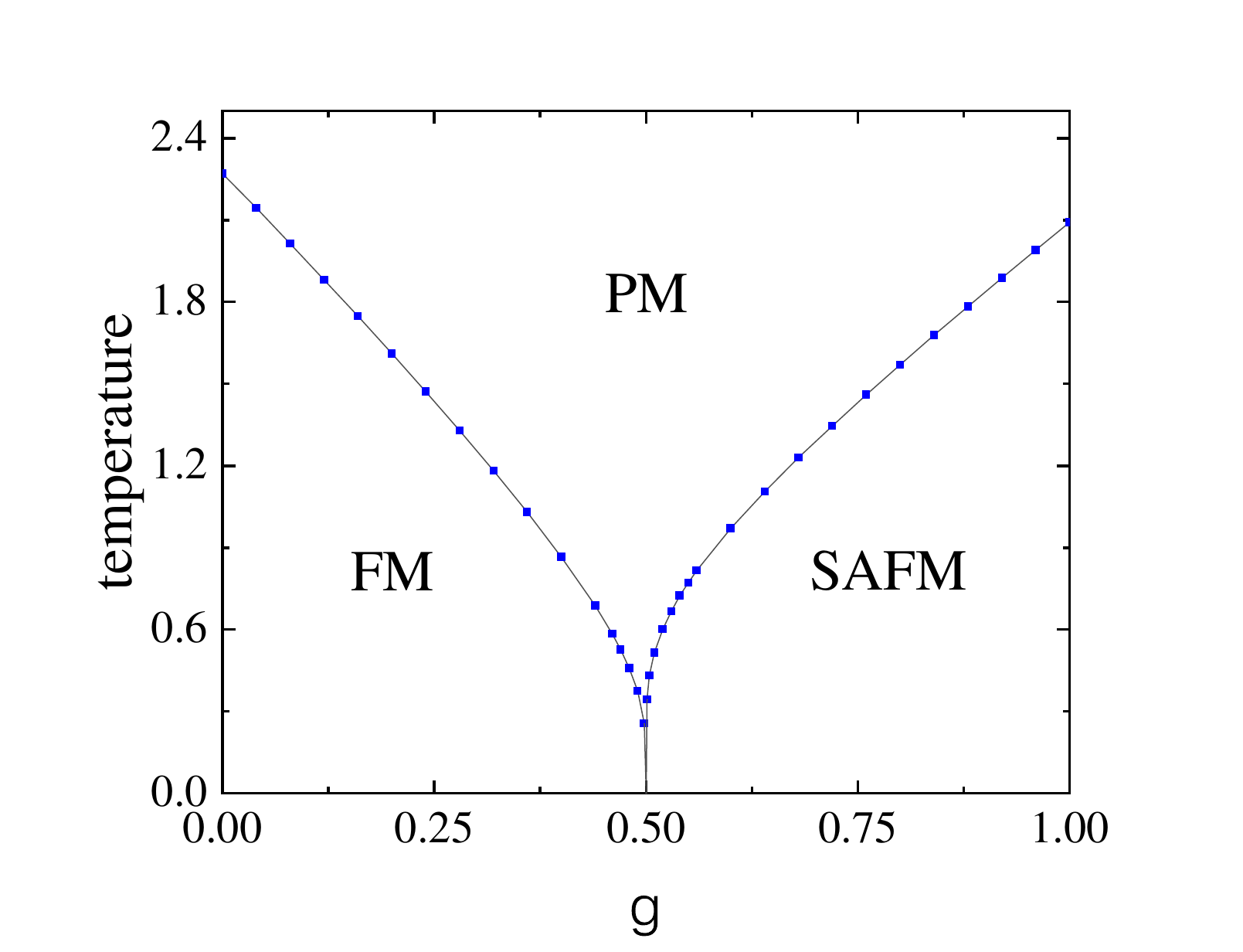}
	\caption{Phase diagram in the parameter space of the temperature and the relative coupling $g$. There are three phases, the ferromagnetic phase (FM), the stripe antiferromagnetic phase (SAFM), and the paramagnetic phase (PM), with phase boundaries indicated by solid lines. The blue squares are data points.}
	\label{fig:Phase-diagram}
\end{figure}

By combining the entanglement entropy with the magnetization obtained from iTEBD, the phase diagram is obtained in Fig.~\ref{fig:Phase-diagram}. It is worth mentioning that the data point at $g=0.5$ is absent. Here $g=0.5$ renders the high degeneracy of the ground state, which invalidates the power method. Thus, we choose the HOTRG for the calculation. However, the physical quantities do not converge with an increasing bond dimension $D$. There were several works claiming that the point at $g=0.5$ has a finite-temperature phase transition~\cite{28, 29, 30}, while other works claimed that the critical temperature is suppressed down to zero due to the high degeneracy, very much like the one-dimensional classical Ising model~\cite{4, 6, 9, 14, 15, 19, huyi}. 

As shown in Fig.~\ref{fig:Phase-diagram}, there are three phases. In the neighborhood of $g=0.5$, the numerical results converge well. Our extensive numerical simulations have been performed for two points $g=0.46$ and $0.55$ on both sides of $g=0.5$;  $g=0.55$ was claimed to be in the region of weak first-order phase transition in Refs.~[\onlinecite{22,24}]. As for $g=0.46$, it had been commonly believed that the phase transition in the whole parameter region of $g<\frac{1}{2}$ is continuous and belongs to the Ising universality class~\cite{10, 15, 19, 20, 27}; however, the analyses from the CMF method~\cite{24}, effective-field theory~\cite{25}, and numerical transfer matrix study~\cite{huyi} suggest that the model may also have a first-order transition line slightly below $g=\frac{1}{2}$.
\begin{figure}[h]
	\centering
	\includegraphics[scale=0.32]{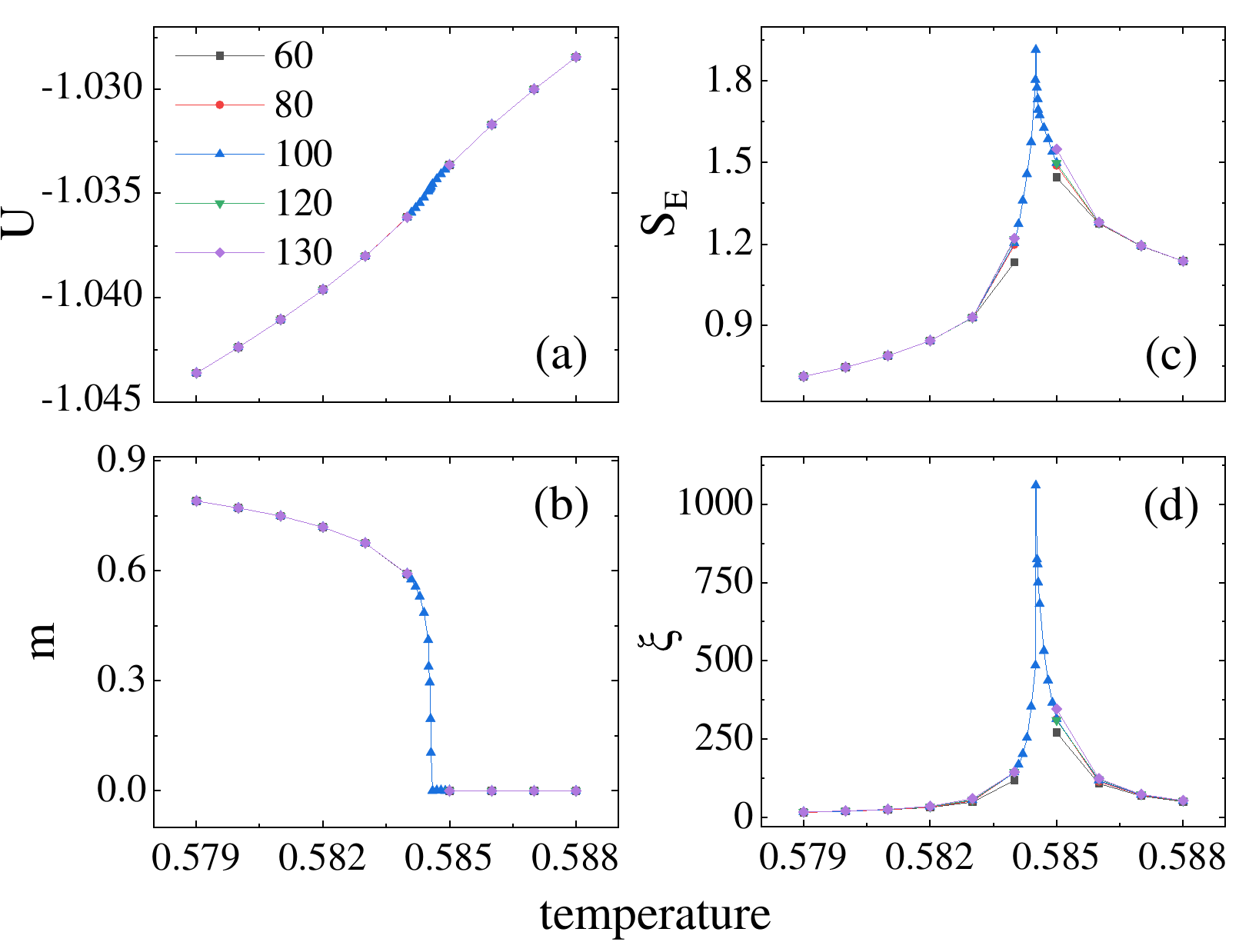}
	\caption{Physical quantities as a function of the temperature at $g=0.46$: (a) internal energy $U$, (b) magnetization $m$, (c) entanglement entropy $S_E$, and (d) correlation length $\xi$. The bond dimension $D$ is $60,80,100,120,130$ in our iTEBD calculations in the framework of the impurity method. }\label{fig:g046}
\end{figure}

\begin{figure}[h]
	\centering
	\includegraphics[scale=0.32]{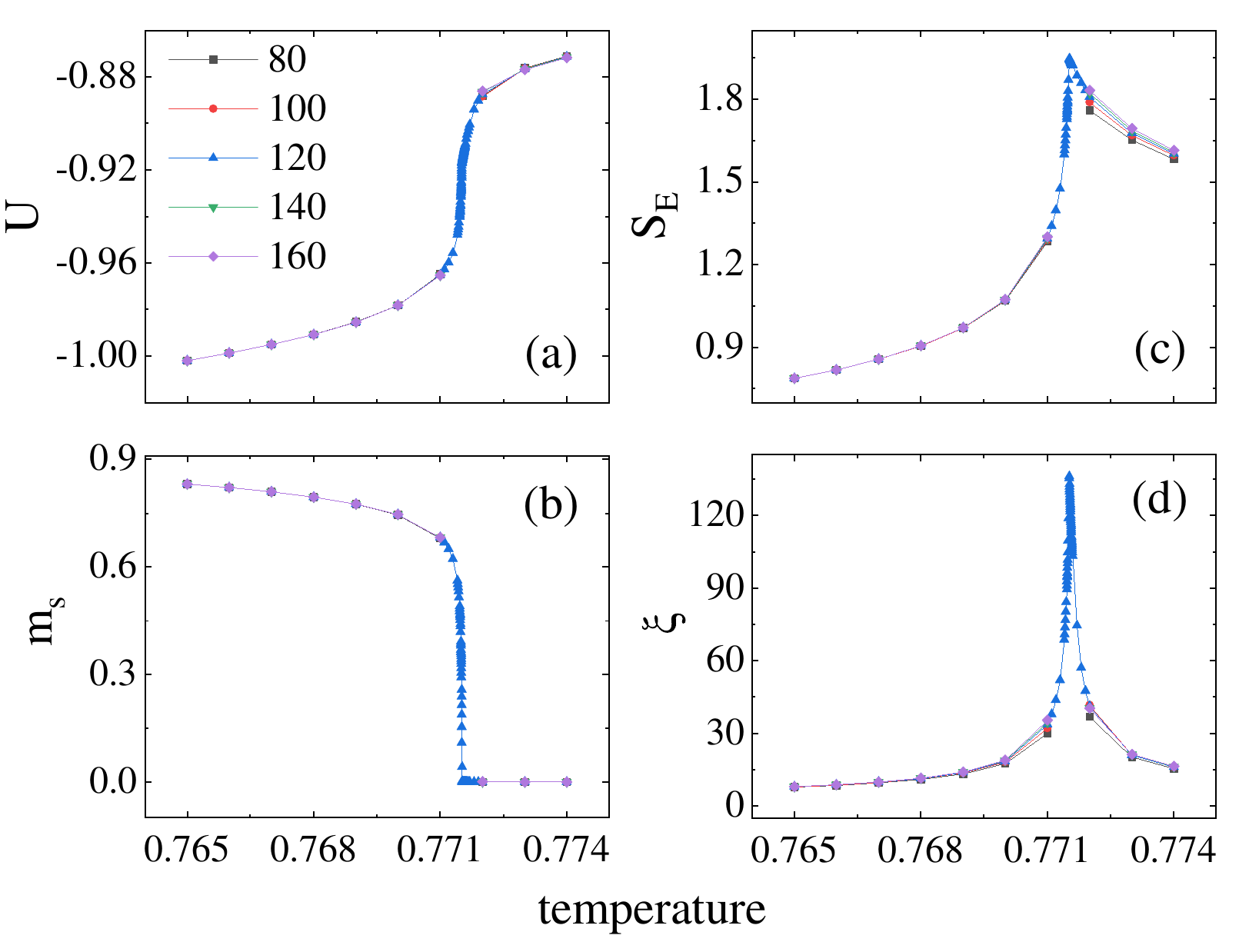}
	\caption{Physical quantities as a function of the temperature at $g=0.55$: (a) internal energy $U$, (b) stripe magnetization $m_s$, (c) entanglement entropy $S_E$ and (d) correlation length $\xi$. The bond dimension $D$ is chosen as $80,100,120,140,160$ in our iTEBD calculations in the framework of the ``impurity" method. }\label{fig:g055}
\end{figure}

As shown in Fig.~\ref{fig:g046}, we compare the results with different bond dimensions with a temperature interval of $10^{-3}$. It seems that there is a jump in the temperature range $(0.584,0.585)$, in which we have used finer data with a temperature interval of $10^{-5}$ at fixed bond dimension $D=100$. Then the jump in the magnetization curve becomes smaller, and there is no discontinuity in the internal energy. As a consequence, the phase transition at $g=0.46$ is of continuous type and belongs to the Ising universality class which we will illustrate below. The recent work using the CMF arrives at the same conclusion~\cite{26}.

Now we move to the case of $g=0.55$. A similar calculation is done, as is shown in Fig.~\ref{fig:g055}. The order parameter is the stripe magnetization $m_s$ with $Z_4$ symmetry breaking. The refined data (the smallest temperature interval is $10^{-6}$) are obtained with bond dimension $D=120$. The MPS entanglement spectrum converges up to $10^{-13}$ at almost all selected temperature points, and up to $10^{-8}$ for only three temperature points, which are extremely close to the transition point. The curves of internal energy $U$ and magnetization $m_s$ are very steep around the critical temperature, which become more prominent in comparison with Fig.~\ref{fig:g046}. However, the discontinuity gradually disappears when denser temperature points are taken. Similar behavior holds at $g = 0.55$. The numerical results suggest that it is also a continuous phase transition. The transition temperature can be read out from the location of the sharp peak in the entanglement entropy and the correlation length.

In particular, we have tested a large bond dimension up to $D=160$ for the convergence check in the case of $g=0.55$. Discerning the phase transition type, especially in between the weak first-order and the continuous types, demands a high precision in numerics. The conclusion that the weak first-order phase transition in the narrow region $\frac{1}{2}<g<g^*$ might be a consequence of indistinguishable continuous change is indicated in physical quantities near the phase transition. 

\subsection{Criticality and universality}
\begin{figure}[h]
	\centering
    \includegraphics[width = 8.5cm]{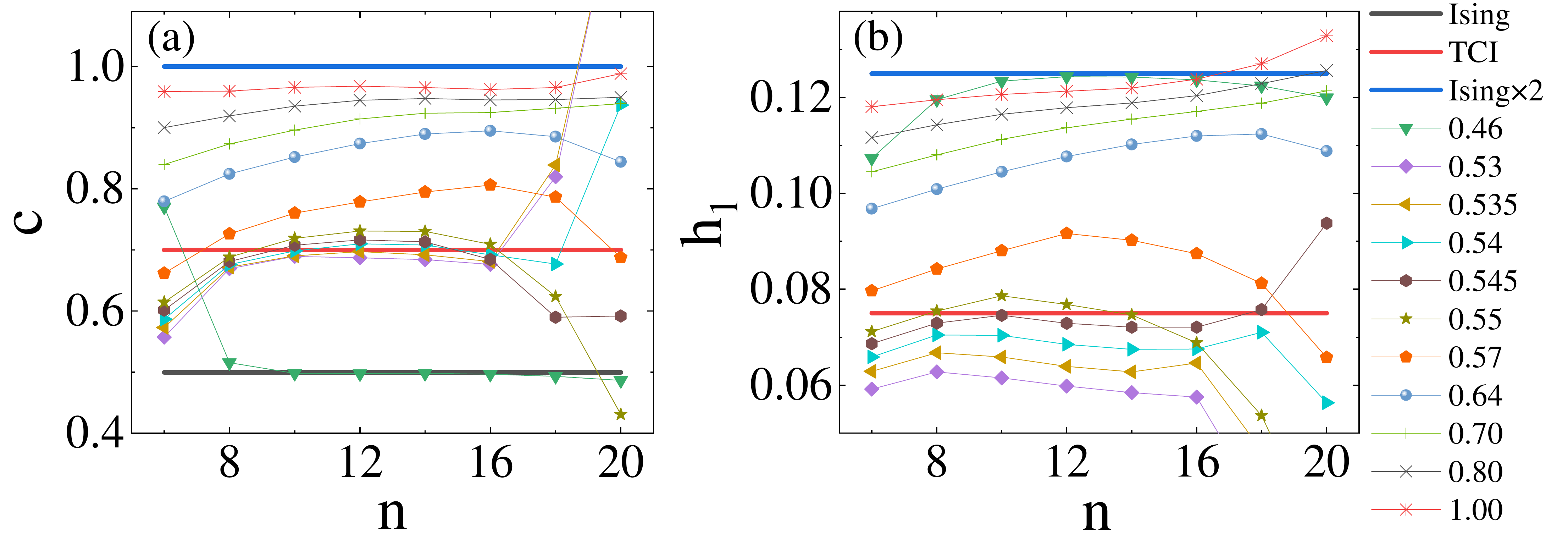}
	\caption{Plot of (a) $c$ and (b) $h_1$ varying with $n$ (RG step) from HOTRG with $D=100$. For reference, three types of universality classes are shown: Ising ($c=0.5$ and $h_1=0.125$), TCI ($c=0.7$ and $h_1=0.075$), and AT ($c=1$ and $h_1=0.125$).}\label{fig:ch1}
\end{figure}

In the scheme of the HOTRG, the critical theory is extracted from the fixed-point tensor in the RG flow. The central charge $c$ and the scaling dimension $h_1$ are obtained by Eq.~(\ref{eq:cri}). 
During the tensor contraction, the size of the system decreases to $\frac{1}{2}$ after one coarse-graining step. After the $n$-th RG iteration, the effective system size in both directions is $L = 2^n$. Correspondingly, the perturbation introduced by the truncation error will gradually destroy the fixed-point tensor when $n$ increases. As a result, $c$ and $h_1$ diverge quickly, as shown in Fig.~\ref{fig:ch1}. 
We observe that the central charge $c$ decreases from about $1$ (the decoupled Ising limit) to $0.7$ and $h_1$ changes to $0.075$ (TCI universality class), as $g$ approaches $g^*$ before complete instability.

The TCI Hamiltonian~\cite{36} can be written as
\begin{equation}
H = -\sum_{\left\langle ij\right\rangle} t_it_j(K+\delta_{\sigma_i\sigma_j})-\mu\sum_{i}t_i.
\end{equation}  
Here $t_i =\sigma_i^2$. There are three options for $\sigma_i$: $0,\pm 1$. For the two spin states $\sigma_i=\pm 1$, $K$ is the coupling constant of a pair with different spin states, and $K+1$ that of a pair with the same spin states. In addition, $\mu$ is the chemical potential for adjusting the average occupation number. Due to one more option of vacancy, there exists a tricritical point belonging to the TCI universality class, where the central charge is given by $c=0.7$~\cite{36}.

It can be observed that as $g$ deviates from 1, the curves of $c$ and $h_1$ varying with RG step become more and more uneven, and the closer $g$ approaches $\frac{1}{2}$, the faster they diverge. This agrees with the discussion above, where the phase transition is accompanied by a significant change in $U$ and $m_s$ when $g$ is close to $\frac{1}{2}$. Thus, it is rather difficult to fix the phase boundary and the fixed-point tensor in the RG flow.

Besides the preceding RG scenario, we also choose a moderate way to contract the tensor network as a cross-check. The alternative way to extract the critical properties is from Eq.~(\ref{eq:lnZTK}). We set $L_x\rightarrow \infty$ with finite $L_y$ ensuring that the requirement $L_x\gg L_y$ in Eq.~(\ref{eq:lnZTK}) holds. The linear fitting of $\text{ln} Z/L_x L_y$ versus $\pi/6{L_y^2}$ works very well in the list of Table \ref{gLyc}, which also indicates a continuous phase transition described by CFT.

The fitted central charge $c$ decreases with decreasing $g$ $(>\frac{1}{2})$, consistent with the result of the HOTRG as demonstrated above. The monotonic behaviors also conform to the $c$ theorem of two-dimensional renormalizable field theory~\cite{Zamolodchikov}. However, the calculation~\cite{21} in the scenario of the transfer matrix does not seem to conform to the $c$ theorem, where $c$ increases when $g$ decreases.

Concerning the consistency with the $c$ theorem in our simulations, two comments are in order. On one hand, starting from decoupled Ising limit, the lattice should be considered in a rotated direction of $\pi/4$, with $J_1$ being treated as a perturbation. We thus use $T_0$ in Fig.~\ref{fig:TensorNcg} as the initial tensor. On the other hand, the usage of very small $L_y$ does not suffice to obtain the conformal data, although there is no truncation error in the numerical simulation. Thus, we adopt the iTEBD scheme to deal with relatively large $L_y$ .

\begin{table}[h]
\caption{The central charge is obtained by fitting $\ln{Z}/L_x L_y \sim \pi/6{L_y^2}$ for selected $L_y$.}
	\begin{tabular} {cccccccccccccccc}
		\hline
		\hline
		$g$  && $c$ &&& $g$  && $c$ &&& $g$  && $c$\\
		\hline
		0.46 &&      0.499 &&& 0.57 &&      0.736 &&& 0.70 && 0.925\\
		0.53 &&      0.631 &&& 0.58 &&      0.760 &&& 0.80 && 0.933\\
		0.55 &&      0.690 &&& 0.60 &&      0.802 &&& 1.00 && 0.978\\ 
		0.56 &&      0.708 &&& 0.64 &&	    0.864 &&& 2.00 && 1.004\\ 
		\hline
		\hline
	\end{tabular}
	\label{gLyc}
\end{table}

\begin{figure}[h]
	\hspace{-0.85cm}
	\centering
	\includegraphics[scale=0.28]{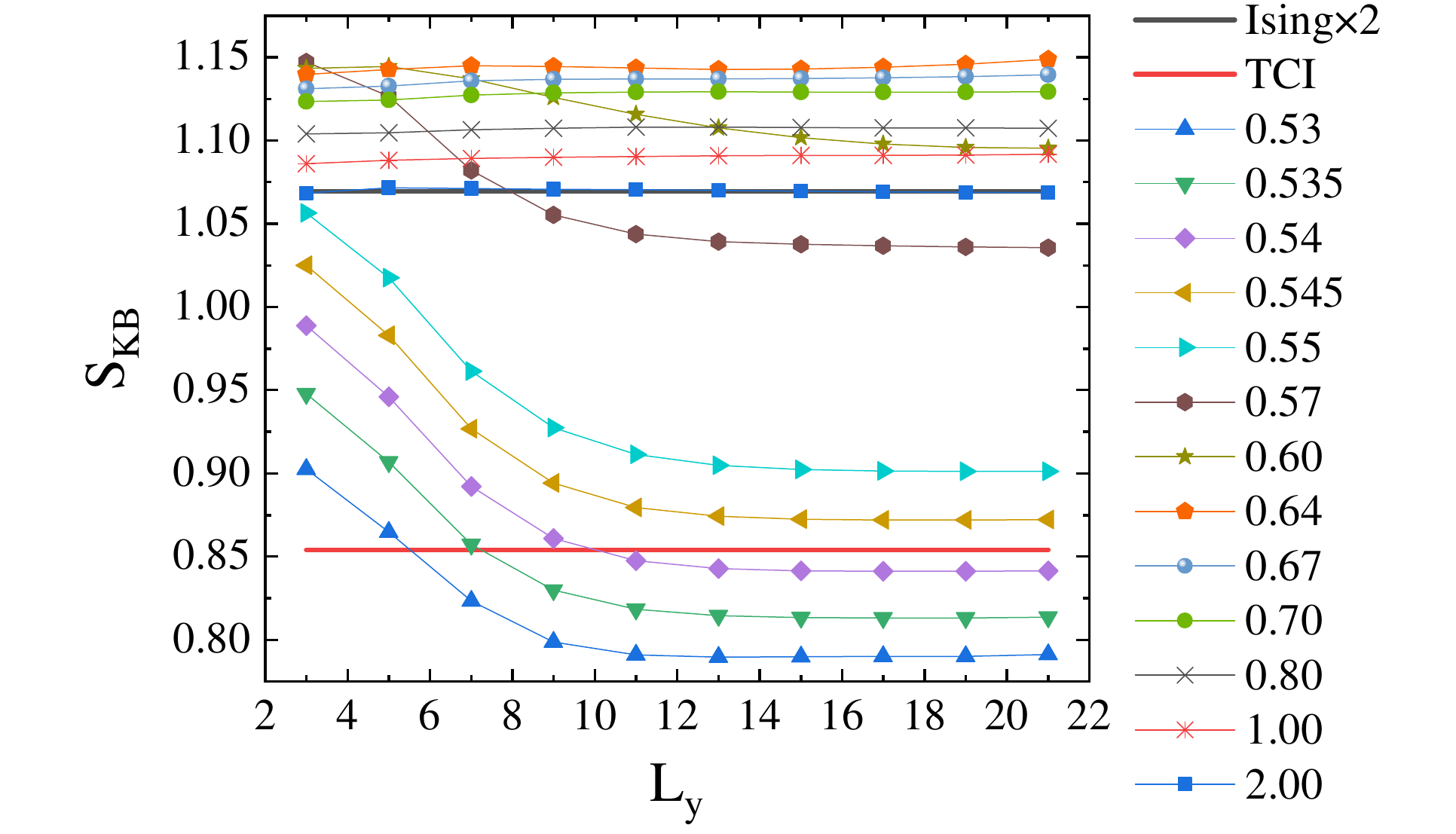}
	\caption{Klein bottle entropy obtained by fitting with different $L_y$ when $g$ takes values in the region $0.5<g\leq2$. The black reference line Ising$\times 2$ is covered by the curve with $g=2.00$.}
	\label{fig:SKB}
\end{figure}
Then we calculate the Klein bottle entropy by fitting with $L_x=100 - 200$, using Eq.~(\ref{eq:lnZTK}). In Fig.~\ref{fig:SKB}, there are two reference lines~\cite{50,36}:
\begin{eqnarray}
S_\text{KB} &=& 2 \times \ln(1+\sqrt{2}/2) \approx 1.0696~~ (\text{Ising} \times 2),\nonumber \\
S_\text{KB} &=& \ln \left[ \frac{(2+\sqrt{2})(s_1 + s_2)}{2\sqrt{s_1^2 + s_2^2}} \right]\approx 0.8543~~ (\text{TCI}).\nonumber
\end{eqnarray}
Here, $s_1 =\sin{(2\pi/5)}$ and $s_2 = \sin{(4\pi/5)}$. In the case of $g=2$, the calculated $S_\text{KB}$ matches the theoretical value of the two decoupled Ising models. In the regime $0.5<g<2$, the curves of $S_\text{KB}$ separate into two bundles. As $g$ decreases from 2, the curves gradually deviate from the Ising$\times2$ line but still remain very even. When $g\lesssim0.64$, a drop in $S_\text{KB}$ arises for small $L_y$; then the curve stabilizes at a value for larger $L_y$. Thereafter, $S_\text{KB}$ decreases as $g$ decreases to about $\frac{1}{2}$ and reaches approximately the value of the TCI universality class at $g^*\approx0.54$.
The combination of the HOTRG and KB entropy suggests the following picture: $J_1$-$J_2$ Ising model evolves continuously from two decoupled Ising models (with  $c = 1$ and $h_1 = 0.125$) to a point described by the TCI universality class (with $c = 0.7$ and $h_1 = 0.075$) as $g$ decreases to $g^*$.

\section{conclusions}
In summary, we have exploited extensive tensor-network calculations to determine the phase diagram and analyze the critical properties of the classical the $J_1$-$J_2$ Ising model. For the cases under debate, we performed detailed simulations. Our numerical results clearly show that the phase transition between ferromagnetic and paramagnetic phases is of the Ising universality class when $g<\frac{1}{2}$. For the phase transition between the SAFM and paramagnetic phases, the results from MC simulations~\cite{22,24,23} and the CVM~\cite{26} indicate that the phase transition is of the continuous type only when $g\gtrsim g^*(\simeq 0.67)$. Our calculation shows that at least in the range of $g\gtrsim 0.54$, the phase transition is continuous, and $g^*\simeq 0.54$ corresponds to the universal class of TCI type. 

The TCI model usually describes the critical region at the tricritical point, at which the continuous Ising phase transition ends. To this point, it could be that the phase transition is of first order in the parameter region $0.5<g\lesssim0.54$. Although tensor network simulations can reach the thermodynamic limit, higher precision is difficult to achieve due to the short-range entanglement near or at criticality in the perspective of the tensor RG flow~\cite{TNR}.
In the regime close to $g=\frac{1}{2}$ demonstrating the criticality, there exist sharp variations in physical quantities, which make it a rather challenging case for the numerical simulations.

Nevertheless, there is another possibility that for all the cases with $g>\frac{1}{2}$, the phase transitions are all continuous. It is generally 
believed that there is no phase transition at $g=0.5$~\cite{4, 6, 9, 14, 15, 19, huyi}. From $g=0.5$ the model might evolve continuously to two decoupled Ising models ($g\rightarrow \infty$) with $c=1$.
Due to the limited bond dimension $D$ and subsequent lack of precision, our simulations become unstable~\cite{Ueno} in the close neighborhood of $g=0.5$, giving rise to the absence of a convincing conclusion. The exploration for the stability and convergence with larger bond dimension is ongoing. 

\section{Acknowledgement}We are indebted to Zhi-Yuan Xie and Hong-Hao Tu for the inspiring discussions. We thank Adil A. Gangat for a critical reading of the manuscript. H. Li was supported by the National Natural Science Foundation of China (Grant No. 11674139, No. 11774420 and No. 11834005), the National R$\&$D Program of China (Grants No. 2016YFA0300503 and No. 2017YFA0302900), and the Research Funds of Renmin University of China (Grants No. 20XNLG19).
L.-P. Y. is supported by National Science Fundation of China, NSFC(Grant No.11874095, and No.12047564), National Science Fundation of Chongqing(Grant No.cstc2018jcyjAX0399), and Research Funds for the Central Universities (No.2018CDXYWU0025).


\begin{thebibliography}{99}
	\bibitem{1}
	L. Onsager, Phys. Rev. {\bf 65}, 117 (1944).
	\bibitem{2}
	M. P. Nightingale, Phys. Lett. {\bf 59}A, 486 (1977).
	
	\bibitem{3}
	R. H. Swendsen and S. Krinsky, Phys. Rev. Lett. {\bf 43}, 177 (1979).
	
	\bibitem{4}
	J. Oitmaa, J. Phys. A:  Math. Gen. {\bf 14}, 1159 (1981).
	
	\bibitem{5}
	K. Binder and D. P. Landau, Phys. Rev. B {\bf 21}, 1941 (1980).
	
	\bibitem{6}
	D. P. Landau, Phys. Rev. B {\bf 21}, 1285 (1980).
	
	\bibitem{7}
	D. P. Landau and K. Binder, Phys. Rev. B {\bf 31}, 5946 (1985).
	
	\bibitem{8}
	J. Oitmaa and M. J. Velgakis, J. Phys. A: Math. Gen. {\bf 20}, 1269 (1987).

        \bibitem{27}
	M. D. Grynberg and B. Tanatar, Phys. Rev. B {\bf 45}, 2876 (1992).

       \bibitem{9}
	J. L. Mor\' an-L\' opez, F. Aguilera-Granja, and J. M. Sanchez, Phys. Rev. B {\bf 48}, 3519 (1993).
	
	\bibitem{10}
	F. Aguilera-Granja, J. L. Mor\' an-L\' opez, J. Phys.: Condens. Matter {\bf 5} A195 (1993). 
	
	\bibitem{11}
	J. L. Mor\' an-L\' opez, F. Aguilera-Granja, and J. M. Sanchez, J. Phys.: Condens. Matter {\bf 6} 9759 (1994).
	
	\bibitem{12}
	E. L\'opez-Sandoval, J. L. Mor\' an-L\' opez, F. Aguilera-Granja, and J. M. Sanchez, Solid State Commun. {\bf 112}, 437 (1999).
	
	\bibitem{13}
	R. A. dos Anjos, J. R. Viana, and J. R. de Sousa, Phys. Lett. A {\bf 372}, 1180 (2008). 
	
	\bibitem{14}
	A. Kalz, A. Honecker, S. Fuchs, and T. Pruschke, Eur. Phys. J. B {\bf 65}, 533 (2008). 
	
	\bibitem{15}
	A. Kalz, A. Honecker, S. Fuchs, and T. Pruschke, J. Phys.: Conf. Ser. {\bf 145}, 012051 (2009).
	
	\bibitem{16}
	N. Alves, Jr. and J. R. Drugowich de Fel\' icio, Mod. Phys. Lett. B {\bf 17} 209 (2003).
	
	\bibitem{17}
	A. Malakisa, P. Kalozoumis, and N. Tyraskis, Eur. Phys. J. B {\bf 50}, 63 (2006).
	
	\bibitem{19}
	S.-Y. Kim, Phys. Rev. E. {\bf 81}. 031120 (2010).
	
	\bibitem{20}
	J. H. Lee, H. S. Song, J. M. Kim, and S.-Y. Kim, J. Stat. Mech. (2010) P03020.
	
	\bibitem{21}
	A. Kalz, A. Honecker, and M. Moliner, Phys. Rev. B {\bf 84}, 174407 (2011).
	
	\bibitem{22}
	S. Jin, A. Sen, and A. W. Sandvik, Phys. Rev. Lett. {\bf 108}, 045702 (2012).

	\bibitem{23}
	A. Kalz and A. Honecker, Phys. Rev. B {\bf 86}, 134410 (2012).

	\bibitem{24}
	S. Jin, A. Sen, W. Guo, and A. W. Sandvik, Phys. Rev. B {\bf 87}, 144406 (2013).

	\bibitem{25}
	A. Bob\' ak, T. Lu\v civjansk\' y, M. Borovsk\' y, and M. \v Zukovi\v c, Phys. Rev. E {\bf 91}, 032145 (2015).

        \bibitem{26}
	N. Kellermann, M. Schmidt, and F. M. Zimmer, Phys. Rev. E {\bf 99}. 012134 (2019).

	\bibitem{huyi}
	Y. Hu and P. Charbonneau, arXiv preprint (2021), arXiv:2106.08442


	\bibitem{28}
	Y. Boughaleb, M. Nouredine, M. Snina,  R. Nassif and M. Bennai, Phys Res. Int. {\bf 2010}, 284231 (2010).
	
	\bibitem{29}
	M. K. Ramazanov, A. K. Murtazaev, and M. A. Magomedov, Solid State Commun. {\bf 233}, 35 (2016).
	
	\bibitem{30}
	R. Timmons and K. De’Bell, Can. J. Phys. {\bf 96}, 912 (2018).

        \bibitem{31}
	A. I. Guerrero, D. A. Stariolo, and N. G. Almarza, Phys. Rev. E {\bf 91}, 052123 (2015).
	
	\bibitem{32}
	A. I. Guerrero and D. A. Stariolo, Physica A {\bf 466}, 596 (2017).
	
	\bibitem{33}
	P. Chandra, P. Coleman, and A. I. Larkin, Phys. Rev. Lett. {\bf 64}, 88 (1990).
	
	\bibitem{34}
	C. Fang, H. Yao, W.-F. Tsai, J. P. Hu, and S. A. Kivelson, Phys. Rev. B {\bf 77}, 224509 (2008).
	
	\bibitem{35}
	R. M. Fernandes, A. V. Chubukov, J. Knolle, I. Eremin, and J. Schmalian, Phys. Rev. B {\bf 85}, 024534 (2012).



	\bibitem{39}
	M. Suzuki, Prog. Theor. Phys. {\bf 51}, 1992 (1974).

	
	
	 \bibitem{adil}
	A. A. Gangat, and Y.-J. Kao, Phys. Rev. B {\bf 100}, 094430 (2019).
         
	\bibitem{SL}
 	Z.-Q. Li, L.-P. Yang, Z. Y. Xie, H.-H. Tu, H.-J. Liao, and T. Xiang, Phys. Rev. E {\bf 101}, 060105(R) (2020). 
 	\bibitem{Levin}
        M. Levin and C. P. Nave, Phys. Rev. Lett. {\bf 99}, 120601 (2007).
        \bibitem{wangc} C. Wang, S.-M. Qin, and H.-J. Zhou, Phys. Rev. B. {\bf 90}, 174201(2014).
    
	\bibitem{46}
	Z. Y. Xie, J. Chen, M. P. Qin, J. W. Zhu, L. P. Yang, and T. Xiang, Phys. Rev. B {\bf 86}, 045139 (2012).

	\bibitem{wangshun}
	S. Wang, Z. -Y. Xie, J. Chen, B. Normand, and T. Xiang, Chin. Phys. Lett. {\bf 31}, 070503 (2014).
	
        \bibitem{44}
	G. Vidal, Phys. Rev. Lett. {\bf 98}, 070201 (2007).
	
	\bibitem{45}
	R. Orus and G. Vidal, Phys. Rev. B {\bf 78}, 155117 (2008).
\bibitem{36}
P. Di Francesco, P. Mathieu, and D. Senechal, \emph{Conformal Field Theory} (Springer, New York, 1997).

     \bibitem{49}
	  H.-H. Tu, Phys. Rev. Lett. {\bf 119}, 261603 (2017).
	  
	  \bibitem{huihai}
	  
     H.-H. Zhao, Z.-Y. Xie, T. Xiang, and M. Imada, Phys. Rev. B {\bf 93}, 125115 (2016)  
      \bibitem{48}
	 Z.-C. Gu and X.-G. Wen, Phys. Rev. B {\bf 80}, 155131 (2009).

\bibitem{affleck} I. Affleck, Phys. Rev. Lett. {\bf 56}, 746 (1986).

\bibitem{HJM} H. W. J. Bl\"ote, J. L. Cardy, and M. P. Nightingale, Phys.
Rev. Lett. {\bf 56}, 742 (1986).
      \bibitem{50}
      L. Chen, H.-X. Wang, L. Wang, and W. Li, Phys. Rev. B {\bf 96}. 174429 (2017).

        \bibitem{Zamolodchikov}
	A. B. Zamolodchikov, JETP Lett. {\bf 43}, 731 (1986).	

        \bibitem{TNR}
	G. Evenbly, and G. Vidal, Phys. Rev. Lett. {\bf 115}, 180405 (2015) ;
	S. Yang, Z.-C. Gu, and X.-G. Wen, Phys. Rev. Lett. {\bf 118}, 110504 (2017).
	\bibitem{Ueno}
	D. Kadoh, Y. Kuramashi, and R. Ueno, Prog. Theor. Exp. Phys. {\bf 2019}, 061B01 (2019) 
	
	
\end{thebibliography}
\end{document}